\documentclass[12pt]{iopart}
\usepackage{graphicx}
\newcommand{\ba}{\begin{eqnarray}}
\newcommand{\ea}{\end{eqnarray}}
\def\be{\begin{equation}}
\def\ee{\end{equation}}
\newcommand{\eqref}[1]{(\ref{#1})}
\begin{document}
\title{Vortex solutions of the Lifshitz-Chern-Simons theory}
\author{N. Grandi}
\address{IFLP, CONICET and Departamento de F\'isica, Facultad de
Ciencias Exactas, Universidad Nacional de La Plata, C.C. 67, 1900 La
Plata, Argentina
\\
Abdus Salam International Centre for
Theoretical Physics, Associate Scheme
\\ Strada Costiera 11, 34151, Trieste, Italy}
\author{I. Salazar Landea}
\address{IFLP, CONICET and Departamento de F\'isica, Facultad de
Ciencias Exactas, Universidad Nacional de La Plata, C.C. 67, 1900 La
Plata, Argentina\\ Abdus Salam International Centre for
Theoretical Physics, ICTP-IAEA Sandwich Training Educational Programme
\\ Strada Costiera 11, 34151, Trieste, Italy}
\author{G.A. Silva}
\address{IFLP, CONICET and Departamento de F\'isica, Facultad de
Ciencias Exactas, Universidad Nacional de La Plata, C.C. 67, 1900 La
Plata, Argentina\\
Abdus Salam International Centre for
Theoretical Physics, Associate Scheme
\\ Strada Costiera 11, 34151, Trieste, Italy}
\begin{abstract}
We study vortex-like solutions to the Lifshitz-Chern-Simons
theory. We find that such solutions exists and have a logarithmically
divergent energy, which suggests that a Kostelitz-Thouless
transition may occur, in which voxtex-antivortex pairs are created
above a critical temperature. Following a suggestion made by Callan
and Wilzcek for the global $U(1)$ scalar field model, we study vortex solutions of the Lifshitz-Chern-Simons
model formulated on the hyperbolic plane, finding that, as expected,
the resulting configurations have finite energy. For completeness, we
also explore Lifshitz-Chern-Simons vortex solutions on the sphere.
\end{abstract}
\maketitle
\section{Introduction}
First proposed in \cite{Kachru}, Lifshitz-Chern-Simons (LCS)
theory can be understood as a modification of the 2+1 dimensional
$z=2$ Lifshitz scalar theory \cite{Ardonne:2003wa},\cite{ghaemi} by the addition of a non-local term. The
action is obtained by dualizing the scalar field into a gauge vector
and then deforming the resulting Lagrangian with a Chern-Simons
three-form. In terms of the original scalar field, the Chern-Simons
term corresponds to a  non-local deformation.

As originally presented, the Lifshitz-Chern-Simons theory models a
system that experiences an isotropic to anisotropic phase transition
at zero temperature. In the anisotropic phase, the $SO(2)$
rotational symmetry enjoyed by the action is spontaneously broken by
the electric field acquiring a non-zero vacuum expectation value.

In \cite{Mulligan:2011eg}, the LCS action was shown to be
equivalent to the model introduced in
\cite{Zhang},\cite{Zhang2} to describe the low energy
behavior of a charged spinless $2d$ fluid in the presence of an
external perpendicular magnetic field. The anisotropic phase appears
through renormalization of the free parameters of the theory, and
reproduces the phenomenology in the presence of a parallel magnetic
field \cite{Xia}.

In the present work, we study vortex solutions of the
Lifshitz-Chern-Simons action. We find that such solutions exist and have
a logarithmically divergent energy. An heuristic argument then
suggest that they may be entropically favored at high temperatures,
leading to a Kosterlitz-Thoules like transition.

It was suggested in \cite{Callan:1989em} that the
logarithmically divergent energy of the global vortex solution to
the charged scalar field model, could be cured by formulating the
model in a negatively curved space. Following such suggestion we
study vortex solutions to the LCS model in the hyperbolic plane. For
completeness we also analyze the vortex solution on the sphere. We
find that the resulting configurations have finite energy. At this
point, one may wonder whether it makes sense to formulate the theory
in curved space. We would like to point out that a scalar theory in
curved space, that reduces to a Lifshitz scalar with $z=2$ on a flat
metric, turns out to describe the off-plane fluctuations of a
tensionless membrane \cite{cit:David}.

The plan of the paper is the following: in section II we present the
Lifshitz-Chern-Simons theory defined on a general curved manifold.
In section III we show that vortex solutions exist in this model by
solving numerically the Euler-Lagrange equations, the relevant
properties of the solution are also discussed in this section. In
section IV we give a brief summary of our results. In the appendix A we review the relation between the Lifshitz-Chern-Simons theory and the Lifshitz scalar one.
\newpage
\section{The Lifshitz-Chern-Simons action}
The Lifshitz-Chern-Simons action in an arbitrary 2-dimensional
curved space with metric $g_{ij}$ is defined as
\ba
S&=&\int \!dt\,  d^2\! x   \sqrt{g}\left(
e_i(\partial_ta_i-\partial_i a_t) -\frac12 \left({\kappa^2}(\nabla_i
e_j)^2+b^2 \phantom{\frac12}\!\!\!\! \right)-
\phantom{\frac12}\!\!\!\!\right.\nonumber\\
&&
\ \ \ \ \ \ \ \
\ \ \ \ \ \ \ \
\left.\phantom{\frac12}\!\!\!\! +{k } \left(a_{t}b
-\frac{\epsilon^{ij}}{2{ \sqrt{g}}}\, a_{i}\partial_{t}a_{j} \right)
-\frac12 \left({m^2}\,(e_i)^2 + \frac{\lambda}{2}
(e_i)^4\right)\right)\,,
\label{LCSD}
\ea
where repeated (or squared) lower latin indexes are understood to be
contracted with the curved inverse metric $g^{ij}$, \emph{i.e.}
$x_iy_i\equiv g^{ij}x_iy_j$. $\nabla_i$ is the standard covariant
derivative, and $b$ is a shorthand for $b=\epsilon^{ij}\partial_i
a_j/\sqrt{g}$ with $\epsilon^{ij}$ the Levi-Civita symbol
$\epsilon^{00}=\epsilon^{11}=0$ and
$\epsilon^{01}=-\epsilon^{10}=1$. The couplings $\kappa^2,k,\lambda$
are dimensionless, while $[m^2]=1/L ^2$. Stated differently, the
Weyl (scale) transformation of the metric $g_{ij}\to\Omega^{2}
g_{ij}$ is a symmetry of the first three parentheses and the
$\lambda$-deformation provided one simultaneously scales the time
coordinates as $t\to\Omega^2 t$, $a_t\to\Omega^{-2}a_t$.

By electric/magnetic duality the first two parentheses are mapped to
the Lifshitz scalar action, while the third one, representing a
Chern-Simons deformation, leads to a non-local term for the Lifshitz
scalar field (see appendix \ref{A}). Finally, the last parentheses in \eqref{LCSD} takes
into account possible deformations of the theory.

The equations of motion derived from the above action read
\ba
&&\label{Gaussd}
\nabla_i e_i + k b= 0\,,
\\&&\label{Faradayd}
\partial_t a_i-\partial_i a_t+\kappa^2 \nabla^2 e_i - \left(m^2  + \lambda (e_j)^2 \right)e_i =0\,,
\\&&\label{Ampered}
g^{ij}\partial_te_j
+ \frac{\epsilon^{ij}}{\sqrt g}\left(\partial_j b+k(\partial_t a_j-\partial_j
a_t)\right)=0\,,
\ea
where $\nabla^2=g^{ij}\nabla_i\nabla_j$. The first equation can be used to obtain the magnetic field $b$ once
the electric field $e_i$ is known.  The second equation on the other
hand defines the electric field in terms of the vector potentials
$a_t,a_j$. Combining the first and second equations to eliminate $b$ and $a_t,a_j$ from the third,
we find
\be
kg^{ij}\partial_te_j = \frac{\epsilon^{ij}}{\sqrt g}\left(\nabla_j \nabla_k
e_k+k^2\left( \kappa^2 \nabla^2 e_j - \left(m^2 + \lambda
(e_k)^2 \right)e_j \right)\right)\,.
\label{eq:single}
\ee
\normalsize
Since we are interested in static solutions, we will drop the time
derivative appearing on the left hand side. In the following
sections we will solve for $e_i$ using \eqref{eq:single} and obtain
$b$ from \eqref{Gaussd}.

For static configurations, the energy functional takes the form
\ba
E&=&\frac12  \int \! d^2\! x \sqrt{g}  \left( {\kappa^2}(\nabla_i
e_j)^2+\frac 1{k^2}(\nabla_i e_i)^2+m^2(e_i)^2 + \frac\lambda{2} (e_i)^4+E_o \right)\,,
\label{venergy}
\ea
where we have added  a zero point (constant) contribution  $E_o$, which will be  adjusted below.

\section{Solutions}
\subsection{Flat space}
In flat space $g_{ij}=\delta_{ij}$ and the   Weyl rescalling mentioned in the previous section
can be realized by the coordinate transformation $x^i\to\Omega x^i$:
the combined transformation $(t,x^i) \to(\Omega^2 t,\Omega x^i)$ and $(a_t,a_i,e_i)\to(\Omega^{-2} a_t,\Omega^{-1} a_i,\Omega^{-1} e_i)$  is a symmetry of the
first three parentheses of action \eqref{LCSD}.
We therefore conclude that the first three parentheses in \eqref{LCSD} describe a critical point with $z=2$ dynamical scaling exponent.

The last parentheses in \eqref{LCSD} amounts
to possible local deformations of the Lifshitz action (see appendix \ref{A} to see its expressions in terms of the Lifshitz scalar $\phi$).
It is immediate to see that the term ${m^2}\,(e_i)^2$ is relevant, dominating in the IR. When $m^2>0$ the $(\partial_ie_j)^2$ can be ignored
and the theory flows to a $z=1$ infrared fixed point. If $m^2<0$, a $(e_i^2)^2$ term with positive coefficient $\lambda$ becomes mandatory to stabilize
the theory. In this situation, a classical expectation value for $e_i$ will appear (see  \cite{frad},\cite{frad2} for related computations involving
the $\lambda$-term).

As discussed above, the order parameter of the theory is a vector on the 2-dimensional plane. We shall classify the
topological disjoint classes of solutions by the winding of the
vector $e_i$ on the circle at infinity. Explicitly, for solutions approaching infinity as $(e_1,e_2)\to (\cos (n\theta),\sin (n\theta))\,e_0$, the topological charge can be defined as
\be
n=\frac{1}{2\pi }\oint_{C}
{\epsilon^{ij}\,\check e_i\,d\check e_j}\,,
\label{wn}
\ee
where $C$ is the circle at infinity and $\check e_i$ is the electric field normalized by its vacuum expectation value. Note that $n$ takes integer values.

\subsubsection{Vacuum solution:}
The true vacuum of the theory corresponds to the lowest energy static
solution of \eqref{Faradayd}-\eqref{Ampered} and  its symmetry will
depend on the sign of $m^2$. For $m^2>0$ the minimum energy solution to
\eqref{eq:single} is $e_i=0$, we call this the ``isotropic'' phase. For $m^2<0$, a uniform vacuum
expectation value develops $e_i = \sqrt{-m^2/\lambda}\ u_i$, with
$u_i$ an arbitrary unit constant vector. This solution breaks the global
$SO(2)$ symmetry enjoyed by the flat space action \eqref{LCSD} and we name it
the ``anisotropic'' phase. It has
 $n=0$ winding number, and vanishing energy provided $E_o=\lambda m^4/2$.

\subsubsection{Vortex solution:}
In what follows we will consider solutions with $n=1$ for the
$m^2=-|m|^2<0$ case and write the flat metric in polar coordinates
as
\be
ds^2 = dr^2 + r^2d\theta^2\,.
\ee
To simplify the equations it is useful to rescale $r=R_o\rho$ where
$R_o=\sqrt{1+k^2\kappa^2}\,\tilde R_o$ with $\tilde R_o=1/|m|k$. The
$n=1$ static vortex ansatz corresponds to
\be
e_\rho(\rho,\theta)=\frac{|m|R_o}{\sqrt{\lambda}}\,f(\rho)\,,\ \ \ \ \ \ \
e_\theta(\rho,\theta)=0.
\label{ansatz}
\ee
Plugging it into
\eqref{eq:single} the resulting equation of motion reads
\be
f'' +\frac1{\rho}f'  -\frac1{\rho^2} f -f \left(f^2 -1\right)=0.
\label{flatvortex}
\ee
Remarkably, this equation coincides with the relativistic $n=1$
global $U(1)$ vortex equation \cite{vilenkin}. The appropriate
boundary conditions for a vortex configuration are
\ba
f(\rho)\to0,&\qquad\,&\rho\to0,\\
f(\rho)\to1,&&\rho\to\infty\,.
\ea
A power series expansion close to the origin shows two independent
behaviors $f(\rho)\sim \rho^{\pm1}$, the linear one being the proper
choice for a non-singular vortex,
\be
f(\rho)=\beta \rho +{\cal O}(\rho^3),\qquad\qquad\qquad\ \ \ \;\rho\ll1\,.
\label{near0}
\ee
At infinity the radial profile asymptotes its vacuum expectation value
as
\be
f(\rho)\approx1-\frac 1{2\rho^2}+{\cal O}\left(\frac1{\rho^{4}}\right),\qquad\qquad\rho\gg1,
\label{asympinfi}
\ee
in this last expression we have dropped  the exponentially decaying homogeneous
piece that arises upon linearizing \eqref{flatvortex} at infinity.

\begin{figure}[htb]
\hspace{22mm}
\includegraphics[width=0.4\textwidth]{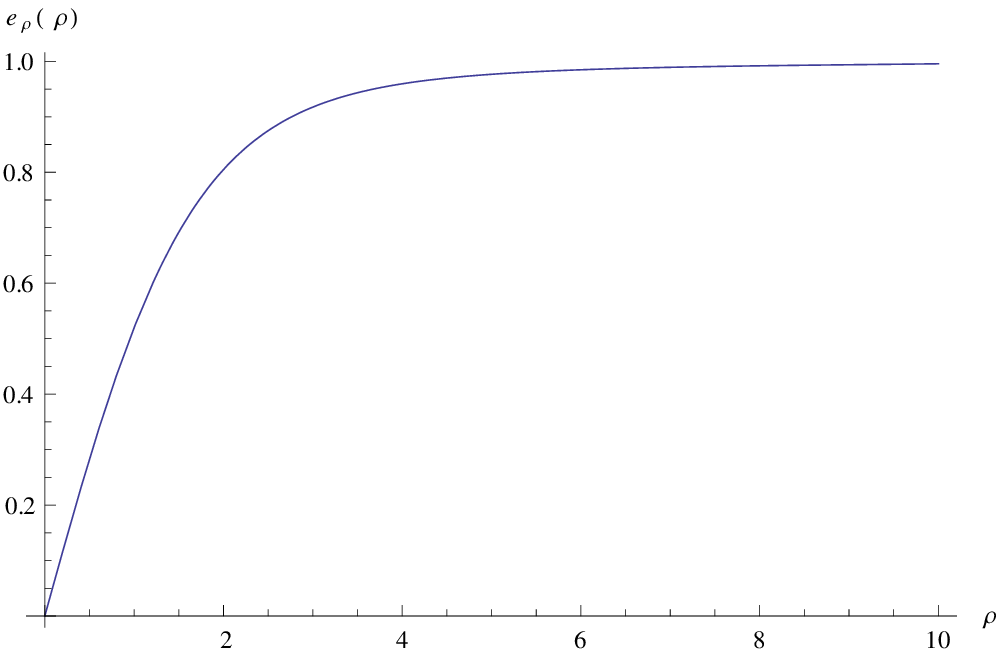}
\hspace{5mm}
\includegraphics[width=0.4\textwidth]{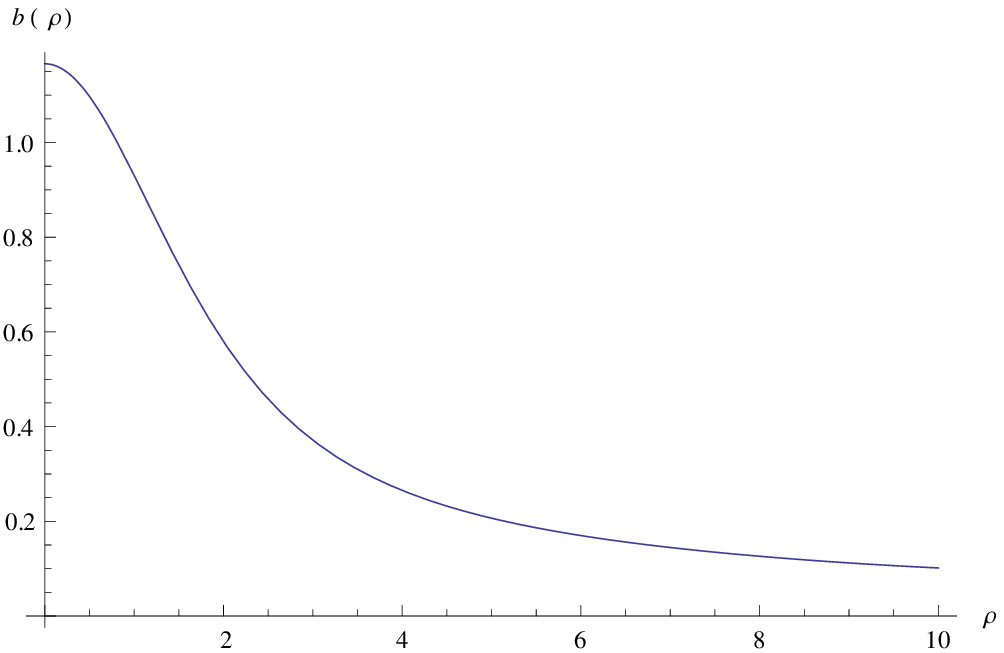}
\caption{Profile of the radial electric and magnetic fields as
funcions of $\rho$: the slope at the origin for $f$ (see
eqn.\eqref{near0}) obtained by our numerical shooting method is
$\beta\approx0.58319$. The electric and
magnetic fields in the plots have been rescaled appropriately.
}
\label{fig}
\end{figure}

We numerically explored the space of solutions of eq.\eqref{flatvortex} shooting from the origin for different values of $\beta$ looking for the asymptotics \eqref{asympinfi}.
The solution we found is plotted in fig.\eqref{fig} with the slope of $f$ at the origin being $\beta\approx0.58319$.

At this point, it is important to stress that the vortex solution for the present model does not exist in the
absence of the Chern-Simons term.
Indeed, a glimpse at equation \eqref{Gaussd} shows that the Chern-Simon coupling $k$ induces a charge density
proportional to the magnetic field $b$, which turns out to be crucial to source the vortex electric field.

\subsubsection{Energy, entropy and free energy:}
In the previous section we have shown the existence of vortex solution in the system.
A natural question to ask is whether the system will choose it as its ground state or not.
In order to answer this question, we write the expression for the energy \eqref{venergy} in the form
\ba
E_v&=& \frac{\pi |m|^4R_o^2}{\lambda}  \int_a^L \!\!  d\rho \rho\left(f'^2+\frac{f^2}{\rho^2}
+\frac12(f^2-1)^2 \right)\,,
\label{venergy2}
\ea
where $a$ is a UV cutoff, and  $L$ the size of the system.
The vortex is regular at the origin, so we do not expect any
singularity in the $a\to0$ limit.
Taking into account the asymptotic behavior \eqref{asympinfi}
one immediately finds that
\be
E_v\approx E_{core}(R)+\frac{\pi |m|^4}{\lambda}\ln\left(\frac LR\right),
\ee
where $E_{core}(R)$ is the finite contribution arising from the energy
density integrated up to a distance $R$ bigger than the core radius $R_o$.

Since the vortex energy diverges logarithmically with the size of the system, the standard Kosterlitz-Thouless argument
follows \cite{Kosterlitz:1972xp}: energetic considerations imply that at negligible temperatures the
system will never choose the vortex configuration, but at finite temperatures, the choice of background
is governed by the Helmholtz  free energy $F=E-T\, S$ and the
entropic contribution could favor vortices for high enough $T$. In our vortex example, this entropic contribution arises from the number of possible places on the plane in which the vortex can sit,
\be
S = 2 k_B
\ln\left(\frac{L}{a}\right)\,.
\ee
The similar logarithmic behavior for both the energy and entropy contributions
will compete on the free energy,
\ba
F &=&  E - T S=
\nonumber\\
&=& E_{core}(R)+\frac{\pi |m|^4}{\lambda}\ln\left(\frac LR\right)- 2k_BT \ln\left(\frac La\right)\,.
\ea
As a result, Kosterlitz and Thouless argued that a temperature should exists above which the system
lowers its free energy by popping vortex-antivortex pairs out of the anisotropic vacuum \cite{Kosterlitz:1972xp},\cite{Kosterlitz:1973xp}.
This phenomenon is described in the literature as vortex unbinding and named topological phase transition.
The critical temperature $T_{KT}$ for this phase transition
can be estimated from the $F=0$ condition, Taking into account the $L$ dependence we find for the present case
\be
T_{KT}\approx\frac{\pi |m|^4}{2k_B\lambda}\,.
\ee
Below this critical temperature the system will be in a quasi-long-range ordered anisotropic phase.
As soon as the system reaches $T_{KT}$ it will be energetically favored to produce vortices and
the quasi-long-range order will be destroyed.

\subsection{Hyperbolic plane}
Negative curvature spaces have been proposed as interesting setups
to cure infrared divergences. The argument is pretty simple
\cite{Callan:1989em}: since the volume of space grows
exponentially with the distance to the origin, Gauss' law implies an
exponential decay for massless fields. In this section we will
analyze the modifications that arise on the vortex solution  when we
formulate the LCS model on the $2d$-hyperbolic plane\cite{nota}.

As in the previous section, the vortex solution is easily found when writting the metric in polar coordinates,
\be
ds^2=R^2(d\rho^2+ \sinh^2\!  \rho \,d\theta^2)\,,
\ee
here $R$ is the curvature radius of the space.
Inserting the static radial ansatz \eqref{ansatz}
into the equations of motion \eqref{eq:single} again with $m^2=-|m|^2$ leads to
\ba
 f'' +\frac1{\tanh \rho}f'-\frac1{\tanh^2\rho}  f -f \left(f^2 -f_o^2 \right) =0\,,
\ea
where $f_o^2=(R^2+\tilde R_o^{2})/R_o^2$. The existence of curvature in the $2$-dimensional space results in an
equation of motion now depending on a parameter $f_o^2$ which cannot be  re-absorbed.

\begin{figure}
\hspace{22mm}
\includegraphics[width=0.4\textwidth]{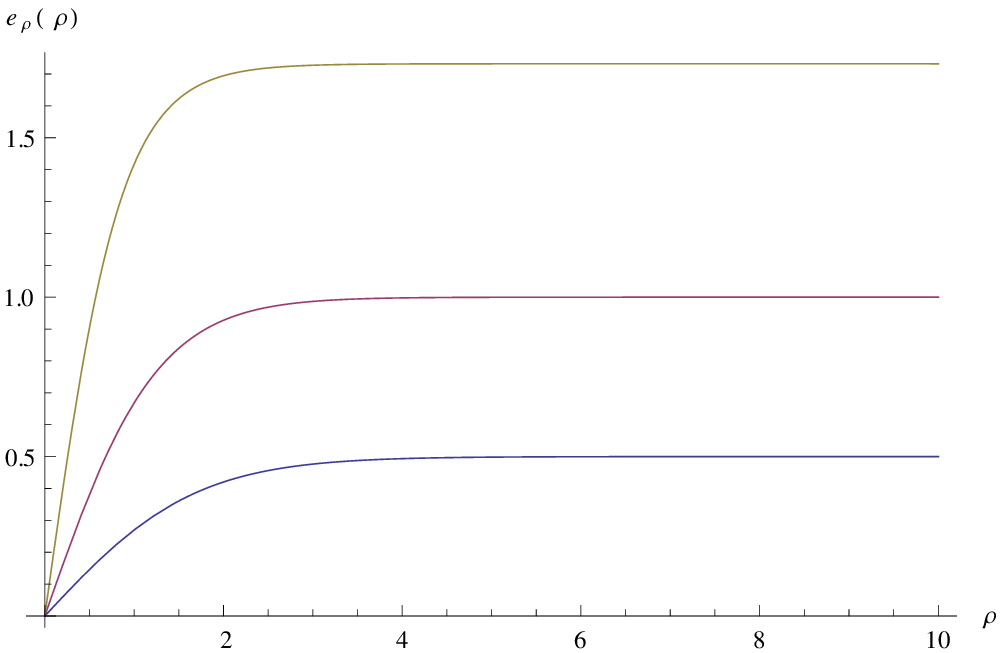}
\hspace{5mm}
\includegraphics[width=0.4\textwidth]{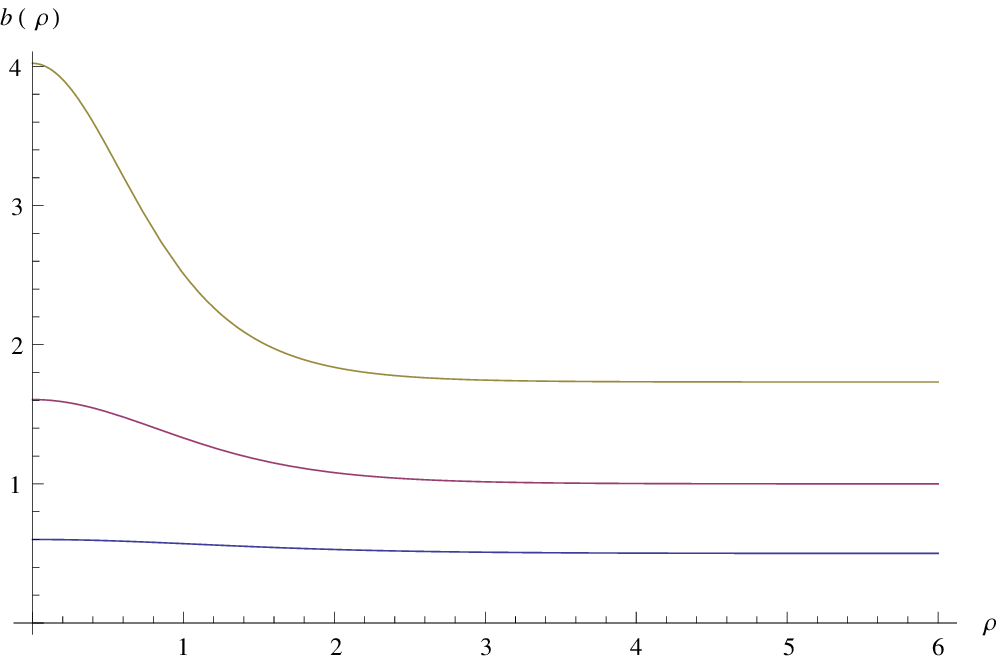}
\caption{Profile of the electric and magnetic fields in the
hyperbolic plane as functions of $\rho$. For $f_o^2=4,2,1.25$ the
corresponding slopes of the electric field at the origin, obtained
by our numerical shooting method, are
$\beta\approx2.01093,0.80291,0.30023$ respectively.}
\label{fig2}
\end{figure}

The behavior for $f$ near the origin coincides with that of flat space  \eqref{near0}, but the large distance
behavior changes drastically. The relevant case for us is $f_o^2>1$ which results in a non-zero value at infinity for the electric field and an
asymptotic  radial profile given by
\be
f(\rho)= \sqrt{f_o^2-1} \left(1-\frac{2}{{f_o^2-2}}e^{-2\rho} +{\cal O}\left(e^{-4\rho}\right) \right)\,,
\label{asymphyp}
\ee
Notice that the exponential decay for the profile is independent of the parameters of the model.
We have depicted in Fig.\eqref{fig2} the numerical solutions we find for different values of $f_o^2$.

The vortex energy functional in the hyperbolic plane case takes the form
\ba
E_v^{^{(hyp)}}
&=&
\frac{\pi |m|^4R_o^4}{\lambda R^2} \int_a^L d\rho  \sinh \rho \left( f' \,\! ^2+\frac{f^2}{\tanh^2\rho}+ 2\frac{ \tilde R_o^2}{ R_o^2}\, \frac{ff'}{\tanh  \rho}
\ +
\right.\nonumber\\&&\left.\ \ \ \ \ \ \ \ \ \ \ \ \ \ \ \ \ \ \ \ \ \ \ \ \ \ \ \ \ \ \ \ \ \ \ \ \ \ \ \ \ \ \ \ \ \ \ \ \
+\
\frac12 {f^4} -\frac{R^2}{R_o^2}f^2+\tilde E_o\right),
\ea
with $a,L$ being respectively UV and IR cutoffs.
The energy density behavior near the origin coincides with that of flat space and
far away from the origin, although the volume element grows exponentially fast at large distances,
the electric field decays more rapidly (see \eqref{asymphyp}) and overwhelms the divergence.
Taking the asymptotic behaviors into account we find that the vortex logarithmic divergence present in
flat space is cured by the spatial negative curvature. We have therefore found that the $n=1$ solution is
a true soliton (finite energy) for the hyperbolic space case.
The appropriate value for $\tilde E_o$ results
\ba
\tilde E_o&=&\frac{(R^2-R_o^2-\tilde R_o^2)(R^2-R_o^2-3 \tilde R_o^2 )}{2 m^4\tilde R_o^4 R_o^4}\,.
\ea

\subsection{Sphere}
\begin{figure}
\hspace{22mm}
\includegraphics[width=0.4\textwidth]{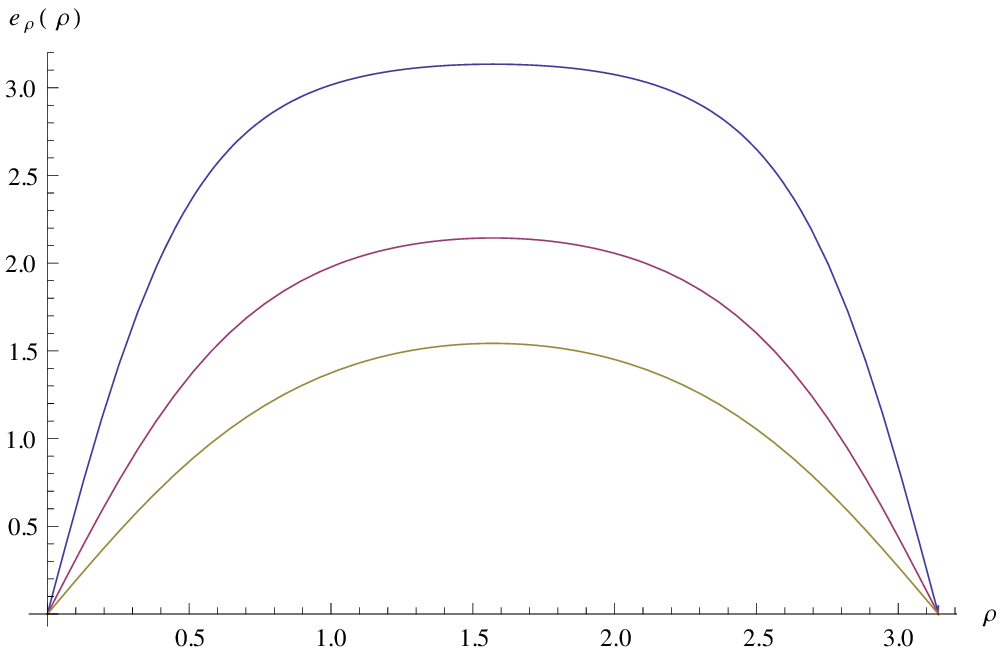}
\hspace{5mm}
\includegraphics[width=0.4\textwidth]{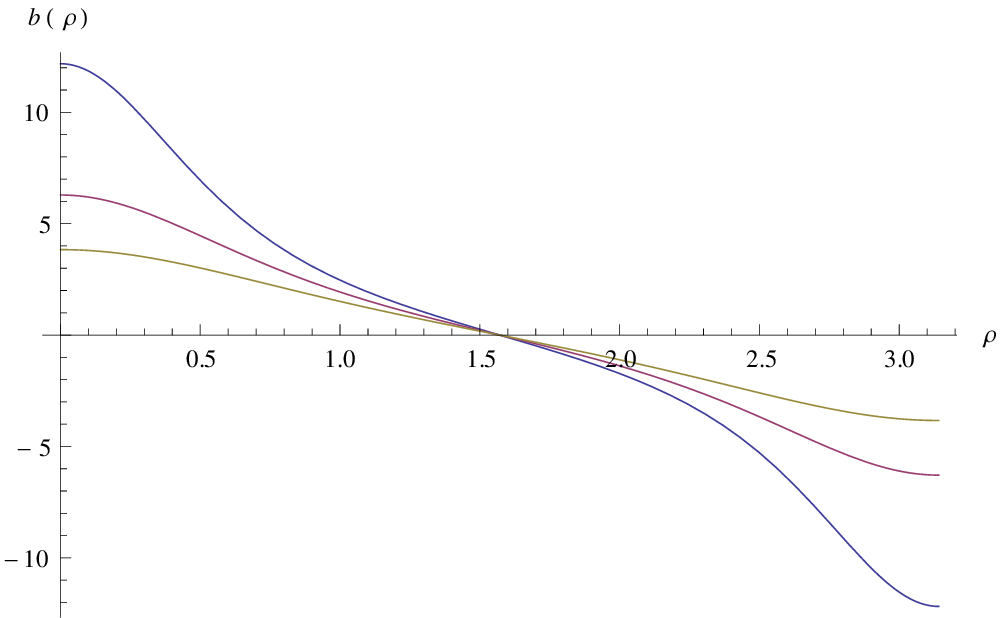}
\caption{Profile of the electric and magnetic fields for the sphere
as function of the azimuthal angle $\rho$ (see eqn.\eqref{spheremetric}). For $f_o^2=3,5,10$, the
slope of the electric field at the origin are
$\beta\approx1.91747,3.143601,6.090353$. }
\label{sphere}
\end{figure}

For completeness we will now consider the LCS model formulated on a sphere with metric
\ba
ds^2=R^2\left(d\rho^2+\sin^2\rho\, d\theta^2\right)\,,
\label{spheremetric}
\ea
here $R$ is the radius of the sphere. Introducing the ansatz \eqref{ansatz}
into \eqref{eq:single}, the equation of motion with $m^2=-|m|^2$  results in
\ba
 f'' +\frac1{\tan \rho}f'-\frac1{\tan^2\rho}  f -f \left(f^2-f_o^2\right) =0\,,
\ea
with $f_o^2=(R^2 -\tilde R_o^{2})/R_o^2$. To avoid singularities, we
should demand the electric field to vanish at the north and south
pole of the $2$-sphere. The appropriate boundary conditions in the
sphere case result
\be
e_\rho(\rho,\theta)|_{\rho=0,\pi}=0\Rightarrow f(0)=f(\pi)=0\,.
\ee
The asymptotic behavior at the poles coincide with the flat space
case and the configurations we obtain with these boundary conditions
correspond to a vortex and an antivortex located at antipodes on the
2-sphere. This last fact is a consequence of \eqref{Gaussd}: no net charge can be
supported on a compact manifold. Therefore the flux lines emanating from the north pole should end on
an equally opposite charge located in our case at the south pole. Some typical profiles  are plotted in Fig.\eqref{sphere}.

The energy for the vortex-antivortex configuration on the 2-sphere takes the form
\ba
E_{v\bar v}^{(sph)}&=&\frac{\pi |m|^4R_o^4}{\lambda R^2} \int_0^\pi d\rho  \sin \rho \left( f' \,\! ^2+\frac{f^2}{\tan^2\rho}+ \frac{2\tilde R_o^2}{R_o^2} \frac{ff'}{\tan  \rho}
+\frac12 {f^4} -\frac{R^2}{R_o^2}f^2\right)\,,\nonumber\\&&~
\ea
In the sphere case the energy of the solution is finite due to the compactness of the
spatial manifold and the regularity of the solution, there
is no need for a compensating constant $E_o$.

\section{Closing Remarks}
We found vortex-like solutions to the Lifshitz-Chern-Simons
theory. The presence of the Chern-Simons term is crucial to the existence of such solutions, since it sources the Gauss law \eqref{Gaussd}. The logarithmically divergent energy suggests that a Kostelitz-Thouless transition may occur on the system. To fully clarify the nature of this phase transition a renormalization group analysis along the lines of \cite{kada} should be performed, this is
currently underway and will be presented elsewhere. Unlike the $U(1)$ global vortex case, extensions to higher winding are non trivial, due to the vector character of the electric field.

Following a suggestion made in \cite{Callan:1989em}, we studied vortex solutions of the Lifshitz-Chern-Simons model formulated on the hyperbolic plane. We found, as expected, that the resulting configurations have finite energy. For completeness, we also explore Lifshitz-Chern-Simons vortex solutions on the sphere. In this last case, as in any compact manifold, the solution we found consisted in a vortex-antivortex pair. An open point is to study the stability of such solution.

An important question is whether the flat space version of action \eqref{LCSD} at $m=0$ is the most general action describing the critical point. In principle, higher order terms like the introduced in \cite{cit:Deser:1999pa} could be added, without breaking the $z=2$ scaling symmetry. We will explore this issue in a separate publication.

\section*{Acknowledgements}
This work has been partially supported by CONICET (PIP2007-0396) and ANPCyT (PICT2007-0849 and PICT2008-1426) grants. We thank ICTP, where part of this work was done, for hospitality. We thank Shamit Karchru and Gustavo Lozano for helpful comments and discussion. We would also like to thank Gerardo Rossini, Horacio Falomir, Daniel Cabra for reading the MSc thesis that originated this work.

\section*{References}

\appendix

\section{Electromagnetic duality in Lifshitz systems}
\label{A}

In this appendix we will shortly review the electromagnetic duality
mapping the $k=0$ action \eqref{LCSD} into a $z=2$ scalar field theory \cite{Kachru},\cite{frad2}.

We start with the flat space action \eqref{LCSD} without the Chern-Simons
deformation,
\small
\ba
 S&=&\int \!dt\,  d^2\! x   \left(
e_i(\partial_ta_i-\partial_i a_t) -\frac12 \left({\kappa^2}(\nabla_i
e_j)^2+b^2 \phantom{\frac12}\!\!\!\! \right) -\frac12
\left({m^2}\,(e_i)^2 + \frac{\lambda}{2} (e_i)^4\right)\right) .
 \nonumber\\
 \label{EMaction}
\ea
\normalsize
The equations of motion following from the action are
\ba
\label{anisoset1aniso}
(\kappa^2\nabla^2-m^2-\lambda e_j^2) e_i= \partial_i a_t- \partial_t a_i
,
\\ \label{anisoset2aniso}
\partial_i e^i=0 ,
\\ \label{anisoset3aniso}
\epsilon_{ij}\partial_jb+\partial_t e_i =0 .
\ea
Written in terms of the gauge invariant fields $e_i,b$ they read
\ba
\label{anisoset1anisox}
(\kappa^2\nabla^2-m^2) \epsilon_{ij}\partial_i e_j-\lambda \epsilon_{ij}\partial_i(e_k^2\, e_j)
= - \partial_t b
,
\\ \label{anisoset2anisox}
\partial_i e^i=0 ,
\\ \label{anisoset3anisox}
\epsilon_{ij}\partial_jb+\partial_t e_i =0 .
\ea

The duality transformation is defined by solving
Gauss law (\ref{anisoset2anisox}) as
\be
e^i=\epsilon^{ij}\partial_j\phi.
\label{df}
\ee
and then  plugging \eqref{df} into equation (\ref{anisoset3anisox}).
This implies that $b$ can be written in terms of $\phi$ as
\ba
b=-\partial_t \phi.
\label{dfb}
\ea
where a time dependent integration ``constant" has been reabsorbed in $\phi$. Replacing \eqref{df}-\eqref{dfb} into \eqref{anisoset1anisox} leads to
\ba
-(\kappa^2\nabla^2-m^2)\nabla^2 \phi+\lambda \partial_i
((\nabla\phi)^2\partial_i\phi)=
\partial_t^2  \phi \label{anisodualaniso}.
\ea
This  equation of motion for the scalar field can be derived from the action
\ba
S[\phi] =\int \!dt\,  d^2\! x
\frac12 \left((\partial_t \phi)^2 - \kappa^2(\nabla^2\phi)^2-
m^2(\nabla\phi)^2 -\frac\lambda2 (\nabla\phi)^4\right)
\ea
The first two terms correspond to the $z=2$ Lifshitz action for a scalar
field. As it happens in its electromagnetic dual, the $m^2$ term drives the theory away from its $z=2$ fixed
point into a $z=1$ IR fixed point. Note that the gauge vector $a_i,a_t$ decouples from $\phi$ and can be integrated out.

In the presence of a CS term, the   dualization described above leads to a non-local action for the
scalar field \cite{Kachru}. An alternative non-local duality transformation leading to a
local equation of motion for the scalar field can be found in \cite{deser}.


\begin{thebibliography}{10}
\bibitem{Kachru}
 M.~Mulligan, C.~Nayak and S.~Kachru,
  ``An Isotropic to Anisotropic Transition in a Fractional Quantum Hall
  State,'',     Phys. Rev. B 82, 085102 (2010),
  arXiv:1004.3570 [cond-mat.str-el].

\bibitem{Ardonne:2003wa}
  E.~Ardonne, P.~Fendley and E.~Fradkin,
  ``Topological order and conformal quantum critical points,''
  Annals Phys.\  {310} (2004) 493,
  arXiv:cond-mat/0311466.

\bibitem{ghaemi}
  P. Ghaemi, A. Vishwanath, T. Senthil,
  ``Finite temperature properties of quantum Lifshitz transitions between valence bond solid phases: An example of `local' quantum criticality,''
  Phys. Rev. B 72, 024420 (2005),
  arXiv:cond-mat/0412409.

\bibitem{Mulligan:2011eg}
  M.~Mulligan, C.~Nayak, S.~Kachru,
  ``Effective Field Theory of Fractional Quantized Hall Nematics,''     Phys. Rev. B 84, 195124 (2011),   arXiv:1104.0256 [cond-mat.str-el].

\bibitem{Zhang}
S. C. Zhang, T. H. Hansson, and S. Kivelson, ``Effective-Field-Theory Model for the Fractional Quantum Hall Effect,''
Phys. Rev. Lett. {62}, 82 (1989)

\bibitem{Zhang2}
S. C. Zhang, ``The Chern-Simons-Landau-Ginzburg theory of the fractional quantum Hall effect,'' Int. J. Mod. Phys. B, (1992)

\bibitem{Xia}
Jing Xia, J.P. Eisenstein, L.N. Pfeiffer, and K.W. West,
``Evidence for a fractional quantum Hall state with anisotropic longitudinal transport,'' arXiv:1109.3219 [cond-mat.str-el].

\bibitem{Callan:1989em}
  C.~G.~Callan, Jr. and F.~Wilczek,
  ``Infrared Behavior At Negative Curvature,''
  Nucl.\ Phys.\ B { 340} (1990) 366.

\bibitem{nota} Local vortices on the hyperbolic plane have appeared in the context of
cylindrical instanton solutions in:  E.~Witten, ``Some Exact Multi - Instanton Solutions of Classical Yang-Mills Theory,'' Phys.\ Rev.\ Lett.\  { 38} (1977) 121.

\bibitem{vilenkin} A. Vilenkin and E.P.S. Shellard, Cosmic Strings and Other Topological Defects (Cambridge Monographs on Mathematical Physics), Ch. 4, CUP, 2000.


\bibitem{Kosterlitz:1972xp}
  J.~M.~Kosterlitz, D.~J.~Thouless,
  ``Long range order and metastability in two dimensional solids
and superfluids,''
  J.\ Phys.\ C { 5 } (1972)  L124.

\bibitem{Kosterlitz:1973xp}
  J.~M.~Kosterlitz, D.~J.~Thouless,
  ``Ordering, metastability and phase transitions in two-dimensional systems,''
  J.\ Phys.\ C { C6 } (1973)  1181-1203.

\bibitem{kada}
  J.~V.~Jose, L.~P.~Kadanoff, S.~Kirkpatrick and D.~R.~Nelson,
  ``Renormalization, vortices, and symmetry breaking perturbations on the two-dimensional planar model,''  Phys.\ Rev.\ B { 16}, 1217 (1977).

\bibitem{cit:Deser:1999pa}
  S.~Deser and R.~Jackiw,
  ``Higher derivative Chern-Simons extensions,''
  Phys.\ Lett.\  B { 451}, 73 (1999)
arXiv:hep-th/9901125.

\bibitem{cit:David} Francois David, ``Geometry and field theory of random surfaces and membranes'', included in D.~Nelson, T.~Piran, S,~Weinberg (Eds) ``Statistical mechanics of membranes and surfaces'', World Scientific (2004) ISBN 981-238-760-9

\bibitem{frad} Eduardo Fradkin, David A. Huse, R. Moessner, V. Oganesyan, S. L. Sondhi, "On bipartite
Rokhsar-Kivelson points and Cantor deconfinement," Phys. Rev. B 69, 224415
(2004), arXiv:cond-mat/0311353.

\bibitem{frad2}A. Vishwanath, L. Balents and T. Senthil, "Quantum Criticality and Deconfinement
in Phase Transitions Between Valence Bond Solids," Phys. Rev. B69 (2004) 224416.

\bibitem{deser}  S.~Deser, R.~Jackiw and S.~Templeton,
  ``Topologically Massive Gauge Theories,''
  Annals Phys.\  {\bf 140} (1982) 372
   [Erratum-ibid.\  {\bf 185} (1988) 406]
   [Annals Phys.\  {\bf 185} (1988) 406]
   [Annals Phys.\  {\bf 281} (2000) 409].

\end{thebibliography}
\end{document}